\documentclass[twocolumn,showpacs,amsmath,aps]{revtex4}
\usepackage{graphicx,color}
\usepackage{bm}
\usepackage[hypertex]{hyperref}
\usepackage{easybmat}

\newcommand{\be}{\begin{equation}}
\newcommand{\ee}{\end{equation}}
\newcommand{\bea}{\begin{eqnarray}}
\newcommand{\eea}{\end{eqnarray}}
\newcommand{\bsube}{\begin{subequations}}
\newcommand{\esube}{\end{subequations}}

\newcommand{\Eq}[1]{Eq.\,(\ref{#1})}

\newcommand{\nl}{\nonumber \\}

\newcommand{\beq}{\begin{equation}}
\newcommand{\eeq}{\end{equation}}
\newcommand{\beqn}{\begin{eqnarray}}
\newcommand{\eeqn}{\end{eqnarray}}
\newcommand{\bsub}{\begin{subequations}}
\newcommand{\esub}{\end{subequations}}

\begin{document}


\title{ Quantum plateau of Andreev reflection
        induced by spin-orbit coupling }

\author{Luting Xu}
\affiliation{Department of Physics, Beijing Normal University,
Beijing 100875, China}
\author{Xin-Qi Li}
\email{lixinqi@bnu.edu.cn}
\affiliation{Department of Physics, Beijing Normal University,
Beijing 100875, China}

\begin{abstract}
In this work we uncover an interesting {\it quantum plateau}
behavior for the Andreev reflection
between a one-dimensional quantum wire and superconductor.
The quantum plateau is achieved by properly tuning the interplay
of the spin-orbit coupling within the quantum wire
and its tunnel coupling to the superconductor.
This plateau behavior is justified to be unique by
excluding possible existences in the cases
associated with multi-channel quantum wire,
the Blonder-Tinkham-Klapwijk continuous model with a barrier,
and lattice system with on-site impurity at the interface.
\end{abstract}


\maketitle

\date{\today}

{\flushleft
Andreev reflection (AR) is a remarkable and useful quantum
coherent process of two particles in correlation,
taking place at the normal metal/superconductor (N/S) interface \cite{And64}.
In this process,
an incident electron in the normal metal picks up another electron
below the Fermi level, forming a Cooper pair across the interface
in the superconductor and leaving a hole in the normal metal \cite{dG66}.
Owing to its versatile applications in probing material properties,
there have been intensive studies on the various AR physics and
related phenomena \cite{Buz05}.
Very restricted examples include AR at the interface of an $s$-
or a $d$-wave superconductor \cite{Dim00,Tana95} and normal systems
of semiconductor \cite{DS99}, ferromagnet \cite{Bee95},
and spintronic material \cite{Tana06}.   }

Of particular interest is involving spin degrees of freedom into
the AR process.
For instance, in the ferromagnetic/superconducting (F/S) hybrid system,
an interplay of the spin degrees of freedom in the ferromagnetic material
not only adds new physics to the AR process,
but has created significant technique of measuring
the spin polarization of magnetic materials \cite{Maz01}.
Another example is the N/S junction
with ``N" a spin-orbit coupling (SOC) system.
It was found \cite{Lv12} that this hybrid system can reveal
the interesting specular AR phenomena predicted
in the graphene-based N/S junction
where the unique band structure plays an essential role \cite{Bee06,Bee08}.

In this paper we present an AR study on the hybrid system
of a quantum wire with Rashba SOC interaction
in contact with an $s$-wave superconductor.
Instead of the popular Blonder-Tinkham-Klapwijk (BTK)
continuous model (approximating the interface as
a $\delta$-function potential barrier) \cite{BTK82},
we perform simulation based on a lattice model.
Remarkably, our simulation reveals an interesting
{\it quantum plateau} behavior for this hybrid system
in one-dimensional (1D) case.
We justify this unique behavior by excluding its existence
in the AR process associated with multi-channel quantum wire, the BTK
continuous model, and 1D lattice system with on-site impurity at the interface.

\vspace{0.2cm}
{\flushleft {\it Model and Methods.}--- }
In this work we consider the hybrid system of a quantum wire
with SOC interaction and in contact with a superconductor.
The quantum wire is modeled as a ribbon in two dimensions,
which is semi-infinite along the longitudinal $x$-direction
and finite in the lateral $z$-direction.
In terms of tight-binding lattice model,
the wire Hamiltonian reads \cite{QF07}
\begin{eqnarray}
H_{\rm w}&=&\sum_i \epsilon_ia_i^{\dag}a_i
-t\sum_i[(a_i^{\dag}a_{i+\delta x}+a_i^{\dag}
a_{i+\delta z})+\rm{H.c}]\nonumber\\
&&+\sum_i[i\alpha(a_i^{\dag}\sigma_xa_{i+\delta z}
-a_i^{\dag}\sigma_za_{i+\delta x})+\rm{H.c}] \,.
\end{eqnarray}
Here we have abbreviated the electron operators
of the $i_{\rm th}$ site with different spin orientation
(in the $\sigma_z$ representation) in  a compact form as
$a_i^{\dag}\equiv (a^{\dag}_{i\uparrow},a^{\dag}_{i\downarrow})$.
$\alpha$ is the SOC coefficient under tight-binding lattice description,
which is related to its counterpart ($\eta$) in continuous model
as $\alpha=\eta/2a$ ($a$ is the lattice constant).
$\epsilon_i$ and $t$ are the tight-binding site energy and
hopping amplitude, while the nearest-neighbor hopping implies
$\delta_x=\delta_z=1$.

For the superconductor we adopt a continuous Hamiltonian,
in momentum $\mathbf{k}$ space which reads \cite{BTK82}
\begin{equation}\label{Hsc}
H_s=\sum_{\mathbf{k},\sigma}\epsilon_{\mathbf{k}}
b^{\dag}_{\mathbf{k}\sigma}b_{\mathbf{k}\sigma}+
\sum_{\mathbf{k}}(\Delta
b^{\dag}_{\mathbf{k}\uparrow}b^{\dag}_{-\mathbf{k}
\downarrow}+\Delta
b_{-\mathbf{k}\downarrow}b_{\mathbf{k}\uparrow}) \,.
\end{equation}
We consider here a two-dimensional (2D) and $s$-wave superconductor.
Then the order-parameter $\Delta$ (assuming real) is independent
of the momentum $\mathbf{k}=(k_x,k_z)$.
The quantum wire and the superconductor are tunnel-coupled,
described as \cite{QF09}
\begin{equation}
H'=\sum_{i,\sigma} \left[ t_ca^{\dag}_{i\sigma}b_{\sigma}(z_i)
   + {\rm H.c.} \right]   \,.
\end{equation}
Here, to reveal the ``nearest-neighbor" coupling feature,
we have converted the (superconductor) electron operator
in momentum space into coordinate representation via
$b_\sigma(z)=\sum_{k_x,k_z}e^{ik_zz}b_{\mathbf{k}\sigma}$.

We attempt to apply the lattice Green's function
technique to compute the Andreev reflection coefficient.
Since the hybrid system under study involves mixing
of electron and hole, and as well their spins, it will be
convenient to implement the lattice Green's function method
in a compact form of the 4-component Nambu representation \cite{QF01}.
In Appendix A we present the particular forms in this representation,
for the quantum wire Hamiltonian
and the superconductor Green's functions (and self-energies).

Moreover, in order to implement the quantum ``transport" approach
based on nonequilibrium Green's function technique
for the interface Andreev reflection problem,
we formally split the (semi-infinite) quantum wire into two parts:
the finite part is treated as ``central device",
and the remaining semi-infinite one as a ``transport lead".
Then, the ``central device" is subject to
self-energy influences from the both (transport) leads.
Based on the surface Green's function technique, the self energy
from the left lead (the SOC quantum wire) is given by \cite{Datta95}
\begin{equation}
\Sigma_L^r(E)=H_{10}g^r(E)H_{01} \, .
\end{equation}
Here, for simplicity, we have dropped the subscript of $H_{\rm w}$.
The surface Green's function $g^r(E)$ can be obtained
as a self-consistent solution from the Dyson equation \cite{Datta95},
$g^r(E)=[E-H_{00}-H_{10}g^r(E)H_{01}]^{-1}$.
In the expressions presented here, we have labeled the first (most-left)
lattice layer of the ``central device" by ``1",
and the most-right layer of the left lead by ``0".
In general, the Hamiltonian matrix elements between them
are still matrices, expanded over the lateral lattice state basis.

Analogously, applying the surface Green's function method,
in Appendix A we carry out the self energy $\Sigma^{r}_{R}$
for the effect of the right lead of superconductor.
Then, the full retarded Green's function of the central device
is given by $G^{r}(E)=[E-H_{\rm w}
-(\Sigma^{r}_{L}+\Sigma^{r}_{R})]^{-1}$,
and the advanced one is its conjugate $G^a(E)=[G^r(E)]^{\dagger}$.
Following the Keldysh nonequilibrium Green's function technique,
a lengthy algebra gives an expression
for the steady-state transport current as \cite{QF09}
\begin{eqnarray}\label{Iss}
I_{ss} &=&\frac{e}{2h}\int
dE ~ \mathrm{Tr}\{[\Gamma_LG^r\Gamma_RG^a]_{ee}(f_L-f_R)\nonumber\\
&&~~~~~~ -[\Gamma_LG^r\Gamma_RG^a]_{hh}(\bar{f}_L-f_R)\nonumber\\
&&~~~~~~ +\Gamma_{Le}G^r_{eh}\Gamma_{Lh}G^a_{he}(f_L-\bar{f}_L)\nonumber\\
&&~~~~~~ +\Gamma_{Lh}G^r_{he}\Gamma_{Le}G^a_{eh}(f_L-\bar{f}_L)\}  \,.
\end{eqnarray}
$f_{L(R)}(E)=f(E-\mu_{L(R)})$ and $\bar{f}_L(E)=f(E+\mu_L)$
are, respectively, the occupied and unoccupied Fermi functions,
with $\mu_{L(R)}$ the chemical potential.
In the above result, ``$e$" and ``$h$" denote the subspace
of electron and hole, which implies the spin
and the lateral lattice states unresolved in explicit basis,
but remaining in a $2Nc\times 2Nc$ matrix form
to be traced after multiplying all the $2Nc\times 2Nc$ matrices.
Finally, the rate matrix $\Gamma_{L(R)}$ in the current
formula is defined from the self energy matrix via
$\Gamma_{L(R)}= i [\Sigma^r_{L(R)}-(\Sigma^r_{L(R)})^{\dagger}]$,
while $\Gamma_{L(R)e}$ and $\Gamma_{L(R)h}$
are their electron and hole blocks.

%

In \Eq{Iss}, the first (second) term describes the electron (hole)
transmission from the left to the right leads,
while the third (fourth) term is for the incidence of
an electron (a hole) accompanied with reflection of
a hole (an electron) to the same (left) lead.
Therefore, for our present interest,
we extract from \Eq{Iss} the AR coefficient as
\bea \label{TAR}
T_{A}(E)={\rm Tr} \left[\Gamma_{Le}(E)G^r_{eh}(E)
\Gamma_{Lh}(E)G^a_{he}(E)\right] \,.
\eea
Note that this formalism has the advantage of allowing for
the incident electron with arbitrary spin orientation
and subject to continuous precession in the ``central device".
The simulated results in this work correspond to arbitrary choice
for the spin orientation of the incident electron.


\begin{figure}
  \centering
  \includegraphics[scale=0.7]{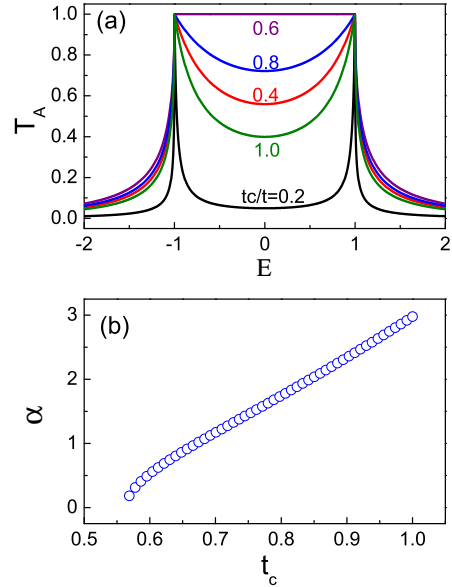}
  \caption{(color online)
  Quantum plateau of AR in 1D quantum wire.
  (a) For $\alpha=0.5 t$, gradual formation of the plateau
  by tuning the contact coupling $t_c$.
  (b) Matching condition between the contact $t_c$
  and the SOC $\alpha$ for the formation of the quantum plateau.  }
\end{figure}

\vspace{0.2cm}
{\flushleft {\it Results and Discussions.}---  }
In our simulations, we use the tight-binding
hopping energy $t$ as the units of all energies,
including $E$, $t_c$, $\alpha$, and $\Delta$.
We commonly set $\Delta=10^{-3}t$
and assume $\epsilon_i=\epsilon_0$ at the Fermi energy.
In Fig.\ 1 we display the central result uncovered in this work
for the 1D quantum wire. First, in Fig.\ 1(a), we visualize
how a quantum plateau of the AR coefficient can appear
by tuning the contact coupling $t_c$ to proper value,
which depends on the SOC $\alpha$ as summarized in Fig.\ 1(b).
In connection with this behavior, we mention that in
the BTK paper \cite{BTK82}, for an 1D wire without SOC,
a similar AR plateau can appear only for vanishing
$\delta$-function potential barrier, which is modeled
to separate the normal and superconducting parts.
In this case, the whole system is a {\it flat} 1D wire,
thus the result seems not so striking, despite the right part
of the wire has suffered the superconducting condensation.
%

In contrast, our system is {\it inhomogeneous}:
the normal part is an 1D wire with SOC;
and the superconducting part has no SOC.
The ``plateau" behavior of the AR coefficient
is thus even more interesting.
The proper matching condition between the SOC $\alpha$ and the
contact coupling $t_c$ for the emergence of the quantum plateau,
as displayed in Fig.\ 1(b), is beyond simple intuition.
When satisfying this matching condition, we have checked that,
by closing the superconducting gap (setting $\Delta=0$)
and remaining all the other parameters unchanged,
the normal transmission coefficient is unity (ideal transmission).
This self consistence provides a support to the AR plateau,
since the AR is anyhow a coherent tunneling process
of two electrons, from the normal part into the superconductor.
However, we remark that in general (the case of unmatched
SOC-$\alpha$ and coupling $t_c$), there is no this sort of
correspondence between the AR and normal transmission coefficients.

\begin{figure}
  \centering
  \includegraphics[scale=0.65]{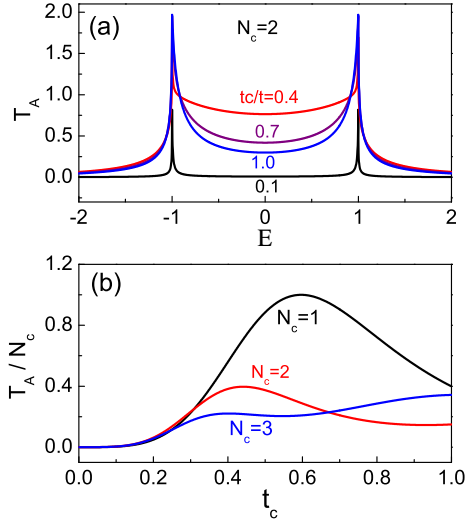}
  \caption{(color online)
  AR coefficient for multichannel quantum wires.
  (a) Results of a two-channel quantum wire where no quantum plateau
  is observed. (b) AR coefficient
  at the Fermi energy by continuously altering $t_c$.
  The results indicate no
  quantum plateau in the multichannel cases (e.g., $N_c=2$ and 3).
  In both (a) and (b) we set $\alpha=0.5t$. }
\end{figure}

\begin{figure}
  \centering
  \includegraphics[scale=0.7]{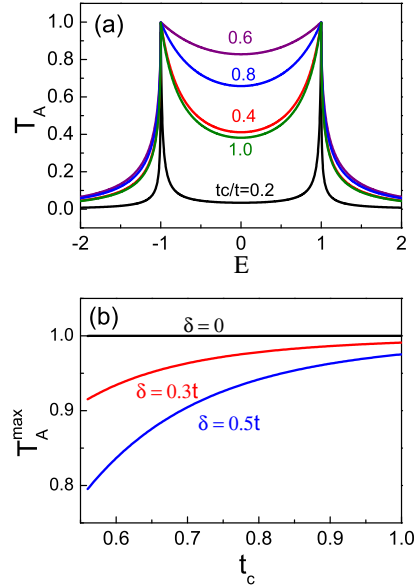}
  \caption{(color online)
  AR coefficient for 1D lattice wire with an impurity
  (with site energy $\epsilon_{\rm im}$) at the interface.
  (a) No quantum plateau formed for nonzero $\delta$
  ($\delta=\epsilon_{\rm im}-\epsilon_0$).
   In this plot we set $\delta=0.5\,t$ and $\alpha=0.5\,t$.
  (b) Maximum AR coefficient ($T^{\rm max}_{A}$) at the Fermi energy
  by optimally matching the SOC $\alpha$ for each $t_c$.
  No quantum plateau appears for nonzero $\delta$.   }
\end{figure}
The quantum plateau behavior is unique,
which we found exists only for 1D SOC quantum wire.
We justify this by simulating multichannel quantum wires,
with results as shown in Fig.\ 2.
In Fig.\ 2(a) and (b), for a given SOC $\alpha$,
by altering the contact coupling $t_c$,
the quantum plateau can no longer be tuned out now.
As a complementary plot, we show in Fig.\ 2(c)
the AR coefficient at the Fermi energy ($E=0$).
For comparative purpose, we rescale the AR coefficient as $T_A/N_c$,
since for the multichannel quantum wire the AR coefficient
(the sum of multiple scattering channels) can exceed unity.
Clearly, we see that, only in the 1D case ($N_c=1$),
can a proper tuning of the contact coupling ($t_c$)
and the SOC $\alpha$ result in the quantum plateau behavior.
In contrast, for multichannel wires (e.g., $N_c=2$ and 3),
the quantum plateau cannot be tuned out,
as demonstrated in Fig.\ 2(c) by noting that
at the edge ($E=\pm \Delta$), the AR coefficient
is ``$N_c$" (the lateral channel numbers).

We further justify the quantum plateau behavior
by considering the BTK 1D continuous model \cite{BTK82}.
The BTK model assumes a normal quantum wire connecting with
a superconductor through a $\delta$-potential barrier (with height $V_0$).
In our case, we further consider the quantum wire
with the Rashba SOC interaction (with strength $\eta$).
In Appendix B, we present a detailed solution for this system
and obtain the AR coefficient as
\begin{eqnarray} \label{TA-cm}
T_A = \frac{(1+x^2)(4Z_2^2+1)}{(4Z_2^2+1)+x^2(2Z_1^2+2Z_2^2+1)^2} \, ,
\end{eqnarray}
where $Z_1=\frac{mV_0}{\hbar^2k_F}$ and $Z_2=\frac{m\eta}{2\hbar^2k_F}$,
with $m$ the electron mass and $\hbar k_F$ the Fermi momentum.
Also, for $E\leq \Delta$,
we have introduced the dimensionless parameter
$x=(\Delta^2-E^2)^{\frac{1}{2}}/E$.
From \Eq{TA-cm}, one can check that
$T_A=1$ only at $E=\Delta$,
and $T_A(E)< 1$ for other $E$.
So we conclude that the quantum plateau behavior does not
appear in the BTK model for nonzero height of barrier.

To understand the above result, which seems in contrast
with the result observed earlier in Fig.\ 1,
let us return to the 1D lattice model.
The $\delta$-potential barrier, in certain sense, is analogous
to an ``impurity" at the end of the 1D lattice chain,
through which the quantum wire is coupled to the superconductor.
Based on this sort of ``impurity" model, we perform further
simulations and present the results in Fig.\ 3.
In Fig.\ 3 (a) we show that,  for a given SOC $\alpha$
and altering the contact coupling ($t_c$),
one can no longer tune out the quantum plateau for the AR coefficient.
Indeed, this differs from what we observed in Fig.\ 1,
but is in consistence with the BTK model discussed above.
In Fig.\ 3 (b) we present a more complete plot
for the absence of the quantum plateau.
For several impurity site-energies ($\epsilon_{\rm im}$),
we display how the quantum plateau behavior disappears.
In this plot, we employ the maximum value ($T^{\rm max}_{A}$)
of the AR coefficient at the Fermi energy,
by optimally tuning the SOC $\alpha$ for each $t_c$,
to illustrate the behavior.

\vspace{0.2cm}
{\flushleft {\it Concluding Remarks.}---  }
We thus arrive at a conclusion that the quantum plateau of AR can be
formed for a homogeneous 1D wire in contact with a superconductor,
as a result of participation of the SOC interaction in the quantum wire.
For this behavior, the SOC effect is essential and not obvious. First,
the incident electron can be initially in arbitrary spin orientation
and experiences continuous spin precession during its propagation.
Second, at the interface, two electrons with opposite spins
coherently enter the superconductor and form a Cooper pair.
But the superconductor is of invariance under spin rotations,
having no unique preferring direction for spin.
This likely leads to an intuition: the AR should not be
affected by the SOC interaction in the quantum wire.
However, our result reveals that the SOC-induced spin precession,
spatially away from the interface,
{\it does} affect the two-electron tunneling into the superconductor
and even a quantum plateau can be induced.
The AR plateau also implies a SOC-induced ``transparency"
for the interface, which does not cause normal reflections.

To summarize, in this work we predict a {\it quantum plateau}
behavior for the Andreev reflection in 1D quantum wire system,
associated with spin-orbit coupling.
It would be of interest to verify this behavior
by experiment in possible engineered 1D systems.


\appendix

\section{ Particulars in Nambu Representation }

The hybrid system under present study involves mixing
of electron and hole, together with their spins.
Let us introduce a generalized Nambu representation \cite{QF01},
$\psi_i = (a_{i\uparrow},a_{i\downarrow},
a^\dag_{i\downarrow},a^\dag_{i\uparrow})^T$,
for the electron operators of the $i_{\rm th}$ layer lattice sites
along the lateral ($z$) direction.
The quantum wire Hamiltonian can be reexpressed
in a compact form as
\begin{equation}
H_{\rm nw}=\frac{1}{2}\sum_i \left[ \psi^\dag_iH_{i,i}\psi_i
+\left(\psi^\dag_iH_{i,i+1}\psi_{i+1}+{\rm H.c.} \right) \right] \,.
\end{equation}
First, the Hamiltonian matrix $H_{i,i+1}$ reads
\begin{eqnarray}
H_{i,i+1}=\begin{bmatrix}
- \widetilde{t}_{+} & 0 & 0 & 0  \\
 0 & - \widetilde{t}_{-} & 0 & 0 \\
 0 & 0 & \widetilde{t}_{+} & 0 \\
 0 & 0 & 0 & \widetilde{t}_{-}
\end{bmatrix}\bigotimes I_{Nc\times Nc}  \,,
\end{eqnarray}
where $\widetilde{t}_{\pm}=t\pm i \alpha$.
The second Hamiltonian matrix, $H_{i,i}$,
has three parts: $H_{i,i}=H_0+H_1+H_2$.
Each is given by, respectively,
\begin{eqnarray}
H_0=\begin{bmatrix}
\epsilon_i&0&0&0\\
0&\epsilon_i&0&0\\
0&0&-\epsilon_i&0\\
0&0&0&-\epsilon_i
\end{bmatrix}\bigotimes I_{Nc\times Nc}
\end{eqnarray}
\begin{eqnarray}
H_1=\begin{bmatrix}
-t&0&0&0\\
0&-t&0&0\\
0&0&t&0\\
0&0&0&t
\end{bmatrix}\bigotimes \begin{bmatrix}
0&1&\\
1&0&\ddots&\text{{\huge{0}}}\\
&\ddots&\ddots&\ddots&\\
&&\ddots&0&1\\
&\text{{\huge{0}}}&&1&0
\end{bmatrix}_{Nc\times Nc}
\end{eqnarray}
\begin{eqnarray}
H_2=\begin{bmatrix}
0&\alpha&0&0\\
\alpha&0&0&0\\
0&0&0&\alpha\\
0&0&\alpha&0
\end{bmatrix}\bigotimes \begin{bmatrix}
0&i&\\
-i&0&\ddots&\text{{\huge{0}}}\\
&\ddots&\ddots&\ddots&\\
&&\ddots&0&i\\
&\text{{\huge{0}}}&&-i&0
\end{bmatrix}_{Nc\times Nc}
\end{eqnarray}

Similarly, for the superconductor (Hamiltonian and Green's functions),
we introduce the 4-component Nambu representation
$\psi^{\dag}_s=(b^\dag_\uparrow,b^\dag_\downarrow,b_\downarrow,b_\uparrow)$.
Originally, the electron operators in the superconductor
Hamiltonian, \Eq{Hsc}, are defined in momentum space.
For the purpose of applying the surface Green's function technique,
we introduce the ``surface" electron operator via
$b_\sigma(z)=\sum_{k_x,k_z}e^{ik_zz}b_{\mathbf{k}\sigma}$.
In this representation, the (retarded) surface Green's function
of the superconductor reads \cite{QF01}
\begin{align}
&g^r_s(z,z',t)=-i\theta(t)\langle\{\psi_s(z,t),\psi^{\dag}_s(z',0)\}
\rangle\nonumber\\
&=-i\theta(t)\times\begin{pmatrix}
\begin{BMAT}{cc.cc}{cc.cc}
\langle\{b_\uparrow,b^{\dag}_\uparrow\}\rangle &
\langle\{b_\uparrow,b^{\dag}_\downarrow\}\rangle &
\langle\{b_\uparrow,b_\downarrow\}\rangle &
\langle\{b_\uparrow,b_\uparrow\}\rangle\\
\langle\{b_\downarrow,b^{\dag}_\uparrow\}\rangle &
\langle\{b_\downarrow,b^{\dag}_\downarrow\}\rangle &
\langle\{b_\downarrow,b_\downarrow\}\rangle &
\langle\{b_\downarrow,b_\uparrow\}\rangle\\
\langle\{b_\downarrow^\dag,b^{\dag}_\uparrow\}\rangle &
\langle\{b_\downarrow^\dag,b^{\dag}_\downarrow\}\rangle &
\langle\{b_\downarrow^\dag,b_\downarrow\}\rangle &
\langle\{b_\downarrow^\dag,b_\uparrow\}\rangle\\
\langle\{b_\uparrow^\dag,b^{\dag}_\uparrow\}\rangle &
\langle\{b^\dag_\uparrow,b^{\dag}_\downarrow\}\rangle &
\langle\{b_\uparrow^\dag,b_\downarrow\}\rangle &
\langle\{b_\uparrow^\dag,b_\uparrow\}\rangle
\end{BMAT}
\end{pmatrix}
\end{align}
Applying the equation-of-motion method,
in frequency domain one obtains \cite{QF09}
\begin{eqnarray}
g^r_S(z,z',E)&=&
-i\pi\rho J_0[k_F(z-z')] \beta(E) \nonumber\\
&& \times \begin{bmatrix}
\sigma_I & (\Delta/E)\,\sigma_z  \\
(\Delta/E) \,\sigma_z  &  \sigma_I
\end{bmatrix}     \,.  \nonumber\\
\end{eqnarray}
In this result, $\sigma_z$ is the Pauli matrix (the third one),
and $\sigma_I$ an identity matrix.
Other notations used here are: the density of states $\rho$,
the Fermi momentum $k_F$,
and the first-type Bessel function $J_0$.
We also introduced:
$\beta(E)=|E|/\sqrt{E^2-\Delta^2}$
for $|E|>\Delta$;
and $\beta(E)=-i E / \sqrt{\Delta^2-E^2}$ for $|E|<\Delta$.

Knowing $g^r_S(z,z',E)$, the self-energy contribution
of the superconductor to the ``central device"
is accordingly obtained via
$\Sigma^r_{R,ij}(E)= |t_c|^2 g^r_S(z_i,z_j,E) $,
where $z_{i(j)}$ corresponds to the ``site"
at the superconductor surface coupled to
the $i_{\rm th}(j_{\rm th})$ site of the quantum wire.

\section{ 1D Continuous Model}


In this Appendix we present a detailed solution for the AR coefficient
based on the BTK 1D continuous model \cite{BTK82} for tunneling through
a $\delta$-function potential barrier (with height $V_0$),
in the presence of Rashba SOC in the quantum wire (with strength $\eta$).
For simplicity but not affecting the conclusion, in the following analysis
we consider only the incident electron with spin-up orientation.
This can reduce the Nambu representation from four- to two-dimensions.
Accordingly, the Bogoliubov-de Gene
Hamiltonian for the total system
is expressed in a compact form as
\begin{eqnarray}
H_{up}=\begin{bmatrix}
H_\uparrow-\mu &\Delta\Theta(x)\\
\Delta\Theta(x)&-H^*_\downarrow+\mu
\end{bmatrix} \,,
\end{eqnarray}
where $\Theta(x)$ is the ``step"-function, and
\begin{subequations}
\begin{eqnarray}
H_{\uparrow}&=&-\frac{\hbar^2}{2m}\frac{d^2}{d^2x}
+i\eta\Theta(-x)\frac{d}{dx}+(V_0-i\frac{\eta}{2})\delta(x) \,, \nl
H_{\downarrow}&=&-\frac{\hbar^2}{2m}\frac{d^2}{d^2x}
-i\eta\Theta(-x)\frac{d}{dx}+(V_0+i\frac{\eta}{2})\delta(x) \,.   \nonumber
\end{eqnarray}
\end{subequations}
In this two-component Nambu representation, the wavefunction is a spinor:
\begin{eqnarray}
\Psi(x,t)=\begin{bmatrix}
f(x,t)\\
g(x,t)
\end{bmatrix} \,.
\end{eqnarray}
For the SOC quantum wire (normal part), substituting the spinor
wavefunction into the Schr\"odinger equation
$i\hbar\frac{d\Psi}{dt}=H_{up}\Psi$,
and considering the stationary solution of
$f(x,t)=u\,e^{iqx-iEt/\hbar}$ and $g(x,t)=v\,e^{iqx-iEt/\hbar}$, we have
\begin{eqnarray}
Eu=\left[\frac{\hbar^2 q^2}{2m}-\hbar q-\mu  \right]u\label{shorda} \,, \nl
Ev=-\left[\frac{\hbar^2 q^2}{2m}-\hbar q-\mu  \right]v\label{shordb} \,.
\end{eqnarray}
Simply, we obtain four spinnor wavefunctions:
\begin{eqnarray}
\Psi^{(e)}_{q^+_j}=\begin{bmatrix}
1\\
0
\end{bmatrix}e^{iq^+_jx} \,,
\end{eqnarray}
and
\begin{eqnarray}
\Psi^{(h)}_{q^-_j}=\begin{bmatrix}
0\\
1
\end{bmatrix}e^{iq^-_jx} \,.
\end{eqnarray}
$q_j^{\pm}$ ($j=1,2$) are given by
$q_j^{\pm}=q_{so}+(-1)^{j-1} \,\tilde{q}^{\pm}$,
with $q_{so}=m\eta/\hbar^2$ and
$\tilde{q}^{\pm}=\sqrt{2m(\mu\pm E+E_{so})}/\hbar$,
where $E_{so}=m\eta^2/(2\hbar^2)$.

Similarly, for the superconductor, the stationary
Schr\"odinger equation reads
\begin{eqnarray}
E\tilde{u}=\left[\frac{\hbar^2k^2}{2m}-\mu  \right]\tilde{u}
+\Delta \tilde{v} \,,\nl
E\tilde{v}=-\left[\frac{\hbar^2k^2}{2m}-\mu  \right]\tilde{v}
+\Delta\tilde{u} \,.
\end{eqnarray}
Here we have assumed $f(x,t)=\tilde{u}\,e^{ikx-iEt/\hbar}$
and $g(x,t)=\tilde{v}\,e^{ikx-iEt/\hbar}$.
Accordingly, we obtain the quasiparticle wavefunctions as
\begin{eqnarray}
\Psi^{(e)}_{\pm k^+}=\begin{bmatrix}
u_0\\
v_0
\end{bmatrix}e^{\pm ik^+x} \,,
\end{eqnarray}
and
\begin{eqnarray}
\Psi^{(h)}_{\pm k^-}=\begin{bmatrix}
v_0\\
u_0
\end{bmatrix}e^{\pm ik^-x} \,,
\end{eqnarray}
where
$u_0^2=1-v^2_0=\frac{1}{2}\left[1+\frac{(E^2-\Delta^2)^{1/2}}{E} \right]$,
and $E=\sqrt{(\hbar^2k^2/2m-\mu)^2+\Delta^2}$
(here taking only the positive root).
In this context, we also applied the following considerations:
for $\hbar^2k^2/2m-\mu>0$, $\tilde u =u_0$ and $\tilde v=v_0$;
while for $\hbar^2k^2/2m-\mu<0$, $\tilde u =v_0$ and $\tilde v=u_0$.
The wavevector numbers read
$k^{\pm}=\sqrt{2m[\mu\pm (E^2-\Delta^2)^{1/2}]}/\hbar$.

As mentioned earlier, we consider incidence of a spin-up electron
with a sub-gap energy.
In this regime, the dominant process is AR.
The associated incident, reflecting, and transmitting waves
are described as
\begin{eqnarray}
\Psi_{i}&=&\begin{bmatrix}
1\\
0
\end{bmatrix}e^{iq^+_1x}  \,, \nl
\Psi_{r}&=&a\begin{bmatrix}
0\\
1
\end{bmatrix}e^{iq^-_1x}+b\begin{bmatrix}
1\\
0
\end{bmatrix}e^{iq^+_2x}   \,,\nl
\Psi_{t}&=&c\begin{bmatrix}
u_0\\
v_0
\end{bmatrix}e^{ik^+x}+d\begin{bmatrix}
v_0\\
u_0
\end{bmatrix}e^{-ik^-x}  \,.
\end{eqnarray}
Following the standard procedures of solving this sort
of tunneling problems, we apply the boundary conditions
at the interface for the wavefunctions and their derivatives.
The first boundary condition reads
\bea
\Psi_S(0)=\Psi_N(0)\equiv\Psi(0) \,.
\eea
Here we have denoted
$\Psi_S=\Psi_{t}$ and $\Psi_N=\Psi_{i}+\Psi_{r}$.
Crossing the $\delta$-function barrier,
the second boundary condition is given by
\begin{eqnarray}
-\frac{\hbar^2}{2m}(\Psi_S^\prime-\Psi_N^\prime)
=(V_0-i\frac{\eta}{2})\Psi(0) \,.
\end{eqnarray}

Noting that $E\leq\Delta<<\mu$,
we can approximate $k^+\simeq k^-\simeq k_F=\sqrt{2m\mu}/\hbar$ and
$q_j^\pm \simeq q_{so}+(-1)^{j-1}\,\sqrt{2m(\mu+E_{so})}/\hbar\equiv q_j$.
We further introduce $\tilde{q}_j= |q_j|/k_F$,
for the sake of brevity in expressions.
More explicitly, the boundary conditions read
\begin{eqnarray}
1+b&=&cu_0+dv_0  \,,\nl
a&=&cv_0+du_0   \,,
\end{eqnarray}
and
\begin{eqnarray}
\frac{i\hbar^2k_F}{2m}(cu_0-dv_0-\tilde{q}_1+b\tilde{q}_2)
&=&(1+b)(V_0-i\frac{\eta}{2}) \,, \nl
\frac{i\hbar^2k_F}{2m}(cv_0-du_0-a\tilde{q}_1)
&=&a(V_0-i\frac{\eta}{2}) \,.
\end{eqnarray}
Solving this set of linear equations yields
\begin{eqnarray}
a&=&\frac{2u_0v_0}{\gamma}(\tilde{q}_1+\tilde{q}_2) \,,\nl
b&=&-\frac{1}{\gamma}(u_0^2-v_0^2)(4Z^2+4iZ\tilde{q}_1
+(\tilde{q}_2-\tilde{q}_1)\tilde{q}_1) \,,\nl
c&=&\frac{u_0}{\gamma}(\tilde{q}_1+\tilde{q}_2)(1+\tilde{q}_1-2iZ) \,,\nl
d&=&\frac{v_0}{\gamma}(\tilde{q}_1+\tilde{q}_2)(1-\tilde{q}_1+2iZ) \,.
\end{eqnarray}
where
\begin{eqnarray}
Z&=&\frac{m(V_0-i\frac{\eta}{2})}{\hbar^2k_F}\equiv Z_1-iZ_2  \,, \nl
Z_1&=&\frac{m V_0}{\hbar^2k_F}=V_0/\hbar v_F  \,,\nl
Z_2&=&\frac{m\eta}{2\hbar^2k_F}=\frac{\tilde{q}_1-\tilde{q}_2}{4} \,,
\end{eqnarray}
and
\begin{eqnarray}
\gamma =(\tilde{q}_1+\tilde{q}_2)+(u_0^2-v_0^2)(4|Z|^2+2) \,.
\end{eqnarray}
Since $E\leq \Delta$, we introduce a real and dimensionless factor
$x\equiv (\Delta^2-E^2)^{\frac{1}{2}}/E$.
Then, $u_0^2=\frac{1}{2}(1+ix)$ and $v_0^2=\frac{1}{2}(1-ix)$.
We finally obtain the AR coefficient as
\begin{eqnarray}
&& T_A(E)= |a|^2 =\frac{(1+x^2)(\tilde{q}_1+\tilde{q}_2)^2}{|\gamma|^2}\nonumber\\
&&~~ = \frac{(1+x^2)(4Z_2^2+1)}{(4Z_2^2+1)+x^2(2Z_1^2+2Z_2^2+1)^2}.  \label{TA}
\end{eqnarray}
We find that at the excitation edge
$T_A(E=\Delta)=1$, otherwise $T_A(E)< 1$.

\vspace{0.1cm}
{\it Acknowledgments.}---
The authors thank Qing-feng Sun for valuable discussions
and help in many technical aspects.
This work was supported by the State ``973"
Project of China (Nos.\ 2011CB808502 \& 2012CB932704)
and the NNSF of China (No.\ 91321106).

\end{document}